\documentclass[twocolumn,showpacs,preprintnumbers,amssymb,aps, prl]{revtex4}
\usepackage{graphicx}
\usepackage{dcolumn}
\usepackage{bm}
\usepackage{latexsym,epsfig}

\usepackage{array}
\usepackage{amstext}

\begin{document}

\title{Parity Nonconservation in Odd-isotopes of Single Trapped Atomic Ions}
\vspace*{0.5cm}

\author{B. K. Sahoo \footnote{bijaya@prl.res.in}}
\affiliation{Theoretical Physics Division, Physical Research Laboratory, Ahmedabad-380009, India}

\author{P. Mandal and M. Mukherjee}
\affiliation{Raman Center for Atomic, Molecular and Optical Sciences, IACS, Kolkata 70032, India}

\date{Recieved date; Accepted date}
\vskip1.0cm

\begin{abstract}
\noindent
We have estimated the size of the light-shifts due to parity nonconservation (PNC) interactions
in different isotopes of Ba$^+$ and Ra$^+$ ions based on the work of Fortson [Phys. Rev. Lett. {\bf 70}, 2383 (1993)]. We have used the nuclear spin independent (NSI) amplitudes calculated earlier by us [Phys. Rev. Lett. {\bf 96}, 163003 (2006); Phys. Rev. A {\bf 78}, 050501(R) (2008)]
and we have employed the third order many-body perturbation theory (MBPT(3))
in this work to estimate the nuclear spin dependent (NSD) amplitudes in these
ions. Ra$^+$ is found to be more favourable than Ba$^+$ for measuring both the NSI
and NSD PNC observables.
\end{abstract}

\maketitle

Parity nonconservation (PNC) in an atom  arises mainly due to 
the exchange of the $Z^0$ boson between the electrons and the nucleus \cite{ginges}.
The interaction leading to such an effect consists of two parts. One of them is
nuclear spin-independent (NSI) and the other is nuclear spin-dependent (NSD). In addition, the interaction 
between the electrons and the nuclear anapole moment (NAM) in an atomic system which is NSD in character,
can also give rise to PNC \cite{ginges,wood}. Wood {\it et. al.} \cite{wood} have reported
the observation of the NAM in atomic Cs.
For an atomic system, the contribution of the NSI and NSD 
components can be extracted by using the addition and 
subtraction of two separate measurements respectively. However, it is not possible to 
distinguish the NSD contributions from the NAM or $Z^0$ exchange from these
measurements, even though the latter is typically larger than the former in heavy atoms. Henceforth in this work, the term NSD interaction will refer to the combination of these two interactions.

Although Wood {\it et. al.} \cite{wood} claim to have observed the NAM in atomic Cs, 
it has so far not been possible to explain this observation from well established 
nuclear data ~\cite{haxton}. Therefore, there is an urgent need to study
the NSD part of atomic PNC.
In atomic transitions when both the NSI and NSD interactons contribute
simultaneously, 
the typical strength of the NSD component is much smaller than its NSI counterpart.
The overall magnitudes of the both interactions grow rapidly with the 
atomic size ~\cite{ginges}. It may therefore appear that heavier atomic systems
would be a natural choice for studying the PNC transitions. But
in a certain cases, the PNC observables can also be enhanced
due to degenracies ~\cite{budker}. To date all the observations of atomic PNC have been reported
in neutral systems \cite{ginges}. However, Fortson in 1993
proposed that a single trapped and laser cooled ion can be used for measuring PNC with an
accuracy that is comparable to that of their neutral counterparts \cite{fortson}. This proposal is based on
the interference of the PNC induced electric dipole amplitude ($E1_{PNC}$) with 
the electric quadrupole (E2) amplitude in the $6s \ ^2S_{1/2} \rightarrow 
5d \ ^2D_{3/2}$ transition in Ba$^+$ which leads to a PNC induced light shift. This approach is being pursued 
at KVI~\cite{wansbeek,versolato} in an
experiment with the ion of the next heavier alkaline earth element Ra$^+$, which
has low-lying optical transitions. This effort has been preceded by many recent
experimental and theoretical studies on properties related to PNC
~\cite{wansbeek,versolato,sherman,koerber,sahoo1,sahoo2,sahoo3}, and it appears that
a high precision result for PNC in these ions is possible. In
addition, a novel idea of observing only the NSD PNC interaction in the
$S_{1/2} \rightarrow D_{5/2}$ transitions of these ions has been reported in
Ref.~\cite{geetha}. 

In this paper, we compare the possible light-shifts due to PNC in the $^2S_{1/2} \rightarrow^2D_{3/2}$ 
and $^2S_{1/2} \rightarrow ^2D_{5/2}$ transitions for  Ba$^+$ and Ra$^+$ 
using our calculated results reported earlier \cite{wansbeek,sahoo1,sahoo2,sahoo3}
as well as from this work. Only the NSD PNC amplitudes are evaluated here
using the third order many-body perturbation theory (MBPT(3)).
The results of these studies will aid in the choice of the hyperfine level 
transitions involving the $S_{1/2}$ and $D_{3/2,5/2}$ states in different isotopes 
for performing PNC experiments in order to extract the NSD PNC observables in a systematic 
and unambiguous way in the two above mentioned ions.

To measure the NSD PNC contribution, we propose to carry out measurements
on $^{135,137}$Ba ($I=3/2$), $^{139}$Ba ($I=7/2$), $^{225}$Ra ($I=1/2$),
$^{223}$Ra ($I=3/2$) and $^{229}$Ra ($I=5/2$). All these isotopes have long
nuclear lifetimes except $^{229}$Ra, where the measurements can be performed in
an on-line facility.

In a single ion experiment, the observable is the PNC electric dipole
transition induced ac-Stark shift \emph{i.e.} the PNC light shift in
the ground state Zeeman sub-levels when a laser drives a $E2$  transition between the
ground state and a meta-stable state. Generalizing the expression given by
Fortson \cite{fortson}, the
measured PNC light shift in the $S_{1/2} \rightarrow D_{3/2,5/2}$ transitions
in a single ion experiment will have a small shift due to PNC as
\begin{eqnarray}
\label{eqn1}
\Delta\omega_{m_F}^{PNC} &\approx& - \frac{Re\Sigma_{m_{F'}}(W_{m_F m_{F'}}^{PNC*}W_{m_F m_{F'}}^{Quad})} {\sqrt{\Sigma_{m_{F'}}|W_{m_F m_{F'}}^{Quad}|^2}},
\end{eqnarray}
and a much larger shift due to the quadrupole coupling
 between the same two levels,
\begin{eqnarray}
\Delta\omega_{m_F}^{Quad} &\approx& \frac{(\omega_0 - \omega)}{2} - \sqrt{\Sigma_{m_{F'}}|W_{m_Fm_{F'}}^{Quad}|^2},
\label{eqn2}
\end{eqnarray}
for a given sub-level $m_F$ of a hyperfine state of angular momentum $F$, where $\omega_0$ 
and $\omega$ are the atomic and optical frequencies, respectively. 
The Rabi frequency for the PNC-induced-dipole transition is given by 
\begin{eqnarray}
W_{m_F m_{F'}}^{PNC} &=& - \frac{1}{2} \sum_{i} (E1_{PNC})_i^{m_F m_{F'}} {\cal E}_i(r=0),
\label{eqn3}
\end{eqnarray}
and the Rabi frequency for the quadrupole transition is given by
\begin{eqnarray}
W_{m_F m_{F'}}^{Quad} &=& - \frac{1}{2} \sum_{i,j} (E2)_{ij}^{m_F m_{F'}} 
\left [ \frac{\partial{ {\cal E}_i(r)}} {\partial{x_j} } \right ]_{r=0},
\label{eqn4}
\end{eqnarray}
where $\cal {E}$ is the applied electric field. We use atomic unit (au) through out this work unless 
mentioned explicitly.

To the best of our knowledge, the $E1_{PNC}$ amplitudes for the
$S_{1/2} \rightarrow D_{3/2}$ transitions due to the NSD interaction for these ions
have not been calculated so far. Preliminary 
calculations of these amplitudes for the $S_{1/2} \rightarrow D_{5/2}$ 
transitions are given in Ref. \cite{geetha} using the relativistic 
configuration interaction (CI) method, but the expression used in that
work for the NSD-interaction is not compatible with our analysis. In what 
follows, we employ a relativistic MBPT(3) method (as described below) here to
evaluate these amplitudes in a systematic fashion for both the NSD PNC transition amplitudes.

The NSD part of the PNC interaction Hamiltonian is given by
\begin{eqnarray}
H_{PNC}^{NSD} &=& \frac {G_F}{2\sqrt{2}} \ {\cal R}_a \ {\bf \vec \alpha} \cdot {\bf \vec I} \ \rho_{nuc}(r) ,
\label{eqn5}
\end{eqnarray}
where $G_F$ is the Fermi constant, $\rho_{nuc}(r)$ is the nuclear potential,
$\vec \alpha$ is the Dirac matrix, ${\bf \vec I}$ is the nuclear spin and
${\cal R}_a$ is a dimensionless constant which has information about NAM.
The $E1_{PNC}$ amplitude due to this Hamiltonian can be written as
\begin{eqnarray}
E1_{PNC} &=& \frac{ \langle \Psi_i^{(0)} | D|\Psi_f^{(1)} \rangle + \langle \Psi_i^{(1)} | D|\Psi_f^{(0)} \rangle} { \sqrt { \langle \Psi_f^{(0)} | \Psi_f^{(0)} \rangle \langle \Psi_i^{(0)} | \Psi_i^{(0)} \rangle}  } ,
\label{eqn6}
\end{eqnarray}
where the subscript $0$ and $1$ represent the atomic wavefunctions and their
first order corrections due to $H_{PNC}^{NSD}$, $i$ and $f$ represent the
valence orbitals in the initial and final
states, respectively, and $D$ is the electric dipole operator.

 In the MBPT method, we define wave operators $\Omega_{v,0}$ and $\Omega_{v,1}$ to calculate the
unperturbed ($|\Psi_v^{(0)}\rangle$) and perturbed ($|\Psi_v^{(1)}\rangle$)
wavefunctions as
\begin{eqnarray}
|\Psi_v^{(k)}\rangle = \Omega_{v,k} |\Phi_v \rangle
&\text{and}& \Omega_{v,k} = \sum_{n=0}^{\infty} \Omega_{v,k}^{(n)} 
\label{eqn7}
\end{eqnarray}
where $ |\Phi_v \rangle$ is the Dirac-Fock (DF) wavefunction obtained
using the Dirac-Coulomb Hamiltonian.

We use the generalized Bloch equation given below to calculate the unperturbed
wavefunctions in our MBPT formulation (for $n \ge 1$) \cite{lindgren}
\begin{eqnarray}
[\Omega_{v,0}^{(n)},H_0] &=& Q V_{r} \Omega_{v,0}^{(n-1)} P - \sum_{m=1}^{n-1} \Omega_{v,0}^{(n-m)} P V_{r} \Omega_{v,0}^{(m-1)} P, \ \ \ \
\label{eqn8}
\end{eqnarray}
where $H_0$ is the DF Hamiltonian, $V_{r}$ is the residual Coulomb
interaction, $\Omega_{v,0}^{(0)}=1$ and $P$ and $Q$ are the projection operators in the model and orthogonal
spaces, respectively; i.e.
\begin{eqnarray}
P = |\Phi_v \rangle \langle \Phi_v |
&\text{and}&  Q = 1 - P.
\label{eqn9}
\end{eqnarray}
Following the similar procedure (for $n \ge 1$), we get
\begin{eqnarray}
[\Omega_{v,1}^{(n)},H_0] &=& Q H_{PNC}^{NSD} \Omega_{v,0}^{(n)} P + Q V_{r} \Omega_{v,1}^{(n-1)} P \nonumber \\
&& \ \ \ - \sum_{m=0}^{n} \Omega_{v,0}^{(n-m)} P H_{PNC}^{NSD} \Omega_{v,0}^{(m)} P \nonumber \\
&& \ \ \ - \sum_{m=0}^{n-1} \Omega_{v,1}^{(n-m-1)} P V_{r} \Omega_{v,0}^{(m)} P \nonumber \\
&& \ \ \ - \sum_{m=1}^{n-1} \Omega_{v,0}^{(n-m)} P V_{r} \Omega_{v,1}^{(m)} P,
\label{eqn10}
\end{eqnarray}
where $\Omega_{v,1}^{(0)}=Q H_{PNC}^{NSD} P$.
For the MBPT(3) approximation, we consider terms up to $n=2$ to evaluate $E1_{PNC}$ given by
\begin{eqnarray}
E1_{PNC} &= & \frac { \langle \Phi_i | \Omega_{i,0}^{\dagger} D \Omega_{f,1} | \Phi_f \rangle
 + \langle \Phi_i | \Omega_{i,1}^{\dagger} D \Omega_{f,0} | \Phi_f \rangle } {norm}  \ \ \ \ \ \
\label{eqn11}
\end{eqnarray}
with $norm= \sqrt { \langle \Phi_f | \Omega_{f,0}^{\dagger} \Omega_{f,0} | \Phi_f \rangle \langle \Phi_i | \Omega_{i,0}^{\dagger} \Omega_{i,0} | \Phi_i \rangle }$.

In the angular momentum relations, we express
\begin{eqnarray}
E1_{PNC} &= & \langle (J_i, I); F_i M_i | D_{eff}^1 + D_{eff}^2 | (J_f, I); F_f M_f \rangle \nonumber \\
    &=& (-1)^{F_i - M_i} \left ( \matrix {F_i & 1 & F_f \cr -M_i & q & M_f } \right ) \mathcal{M},
\label{eqn12}
\end{eqnarray}
where $F$ is the total angular momentum due to the electron angular
momentum ($J$) and the nuclear spin ($I$) with its azimuthal component $M$. Here
$q=-1,0,1$ depending upon the values of $M_i$ and $M_f$. $\mathcal{M}=\langle (J_i, I); F_i || D_{eff}^1 + D_{eff}^2 || (J_f, I); F_f \rangle$ is the reduced
matrix element of effective rank one operators with
\begin{eqnarray}
D_{eff}^1 = \frac{\Omega_{i,0}^{\dagger} D \Omega_{f,1}} {norm}  \ \ \ \
\text{and} \ \ \ D_{eff}^2= \frac{\Omega_{i,1}^{\dagger} D \Omega_{f,0}} {norm}.
\label{eqn13}
\end{eqnarray}
The above expression is non-zero for $F_i = F_f, F_f \pm 1$.

With the expansion $H_{PNC}^{NSD}= \sum_{\mu} (-1)^{\mu} I_{\mu} K^{-\mu}$, it gives
\begin{widetext}
\begin{eqnarray}
\langle (J_i, I); F_i || D_{eff}^1 || (J_f, I); F_f \rangle &=& \eta
\sum_{j\ne i} (-1)^{j_i - j_f +1} \left \{ \matrix {F_f & F_i & 1 \cr J_j & J_f & I } \right \} \left \{ \matrix {I & I & 1 \cr J_j & J_i & F_i } \right \} \frac{ \langle J_f || D || J_j \rangle \langle J_j || K^{1} || J_f \rangle}{ \epsilon_i - \epsilon_j}
\label{eqn14}
\end{eqnarray}
and
\begin{eqnarray}
\langle (J_i, I); F_i || D_{eff}^2 || (J_f, I); F_f \rangle &=& \eta
\sum_{j\ne f} (-1)^{F_i - F_f +1} \left \{ \matrix {F_f & F_i & 1 \cr J_i & J_j & I } \right \} \left \{ \matrix {I & I & 1 \cr J_j & J_f & F_f } \right \} \frac{ \langle J_f || K^{1} || J_j \rangle \langle J_j || D || J_f \rangle}{ \epsilon_f - \epsilon_j},
\label{eqn15}
\end{eqnarray}
where $\eta=\sqrt{I(I+1)(2I+1) (2F_i+1) (2F_f+1)}/norm$, $\epsilon_i$ represents
orbital energy for $i$ and the matrix element in terms of single particle
orbitals is given by
\begin{eqnarray}
\langle J_i || K^{1} || J_f \rangle = i \frac{G_F}{2 \sqrt{2}} \frac{\kappa_a}{I} \int_0^{\infty} dr \rho_{nuc}(r) [\langle \kappa_i || \sigma || -\kappa_f \rangle P_i(r) Q_f (r) - \langle - \kappa_i || \sigma || \kappa_f \rangle Q_i(r) P_f (r) ],
\label{eqn16}
\end{eqnarray}
\end{widetext}
for $P(r)$ and $Q(r)$ being the large and small radial components of Dirac
wavefunction and $\sigma$ is the Pauli spinor with $\langle \kappa_i || \sigma || \kappa_f \rangle = (\kappa_i + \kappa_f -1) \langle \kappa_i || C^1 || \kappa_f \rangle$
for the Racah tensor $C$ and the relativistic quantum number $\kappa$.

\begin{table*}[t]
\caption{Induced light-shifts due to E2 and PNC interactions in different 
isotopes of Ba$^+$ and Ra$^+$. ${\cal R}_a \approx 0.2$ from the Cs data 
\cite{wood,haxton} and electric fields as
$2\times10^6$~V/m are considered for the practical estimations.}
\begin{center}
\begin{tabular}{lccccccccc}
\hline \hline
Transition & $\langle j_i,1/2 | E1_{PNC}^{NSI} | j_f, 1/2 \rangle$ & $\langle j_i || E2 || j_f \rangle$ & $F_i$ & $F_f$ & $\langle F_i || E1_{PNC}^{NSD} || F_f \rangle$ & $m_F$ & $\Delta\omega^{Quad}/2\pi$ & $\Delta\omega_{PNC}^{NSI}/2\pi$ & $\Delta\omega_{PNC}^{NSD}/2\pi$ \\
 $j_i \rightarrow j_f$  &  ($\times 10^{-11}iea_0$) \cite{sahoo1,wansbeek} &  (au) \cite{sahoo2,sahoo3}  &     &       &  ($\times 10^{-14}iea_0/{\cal R}_a$)  &    &      (MHz)                     &             (Hz) & $\times10^{-4}$(Hz) \\
\hline
 & & \\
\multicolumn{6}{l}{$^{135 / 137}$Ba$^+$ ($I=3/2$)}  & &  \\
6s $\rightarrow$ 5d$_{3/2}$ & 2.46 & 12.74 & 2  & 3 & 97.18 & 1 & $7.62$  & $-0.24$ & $6.08$ \\
                            & &  & 1 & 2 & $-89.40$ & 1 & $9.53$  & $-0.37$ & $-14.03$ \\
6s $\rightarrow$ 5d$_{5/2}$ & 0 & 15.96 & 2  & 3 & 1.37 & 1 & $11.13$ & 0 & $0.09$ \\
\multicolumn{6}{l}{$^{139}$Ba$^+$ ($I=7/2$)} & &  \\
6s $\rightarrow$ 5d$_{3/2}$ & 2.46 & 12.74 & 3 & 3 & 103.57 & 1 & $5.07$  & 0.42 & $12.29$ \\
                            & & & 3  & 2 & $-105.55$ & 3 & $3.6$  & $-0.47$ & $-14.75$ \\
                            & &  &  & &  & 1 & $3.6$  & $-0.1$ & $2.95$ \\
6s $\rightarrow$ 5d$_{5/2}$ & 0 & 15.96 & 3  & 2 & $0.608$ & 3 & $12.87$  & 0 & $0.08$ \\
                            & &  &  & &  & 1 & $12.87$  & 0 & $-0.02$ \\
\multicolumn{6}{l}{$^{225}$Ra$^+$ ($I=1/2$)} & & \\
7s $\rightarrow$ 6d$_{3/2}$ & 46.4 & $-14.87$ & 1  & 2 & 991.75 & 1 & $38.95$  & 9.97 & $155.6$ \\
\multicolumn{6}{l}{$^{223}$Ra$^+$ ($I=3/2$)} & & \\
7s $\rightarrow$ 6d$_{3/2}$ & 46.4 & $-14.87$ & 2  & 3 & 1173.45 & 1 & $22.03$  & $-4.7$ & $73.35$ \\
                              & &  & 1 & 2 & $-1017.49$ & 1 & $27.54$  & $-7.05$ & $-159.64$ \\
7s $\rightarrow$ 6d$_{5/2}$ & 0 & $-19.04$ & 2  & 3 & $-17.52$ & 1 & $32.08$ & 0 & $-1.1$ \\
\multicolumn{6}{l}{$^{229}$Ra$^+$ ($I=5/2$)} & & \\
7s $\rightarrow$ 6d$_{3/2}$ & 46.4 & $-14.87$ & 3  & 2 & 325.94 & 3 & $29.44$  & $-4.19$ & $45.56$ \\
                            & &  &  & & & 1 & $29.44$   & 0.83 & $-9.11$ \\
                            & &  & 2 & 2 & $1147.73$ & 1 & $17.42$  & 7.42 & 152.18 \\
7s $\rightarrow$ 6d$_{5/2}$ & 0 & $-19.04$ & 3  & 2 & $-5.12$ & 3 & $17.5$  & 0 & $-0.72$ \\
                            & &  &  & &  & 1 & $17.5$  & 0 & $0.14$ \\
\hline \hline
\end{tabular}
\end{center}
\label{tab1}
 \end{table*}

In the method proposed by Fortson, the PNC electric dipole
transition induced light shift is observed as a frequency shift of
the ground state Larmor frequency in spin zero isotopes. Any
fluctuation of the quadrupole transition induced light shift does not
modify the uncertainty of the PNC light shift measurement, since it cancels 
out in the Zeeman transition due to its $m_F$ dependence (see Eqs.~(\ref{eqn1}) 
and (\ref{eqn2})). However, in the case of non-zero spin isotopes of these ions, 
the $\pm m_F$ transitions are forbidden by the $E2$ selection rules. 

The estimated magnitudes of the PNC light shifts for both NSI and NSD in the
$S_{1/2}\rightarrow D_{3/2,5/2}$ transitions are given in Table ~\ref{tab1}
(only non negligible values are given). For the 
calculations, we have used our earlier results for NSI PNC and electric quadrupole transition
amplitudes \cite{wansbeek,sahoo1,sahoo2,sahoo3} 
and the NSD amplitudes have been evaluated in this work. On practical grounds, we have taken
the strength of the electric field as 
$2\times10^6$~V/m; this value optimizes the quenching rate for Ba$^+$, and 
$R_a\approx0.2$, obtained from the Cs NSD PNC studies \cite{wood,haxton}.

It is clear from the above table, the NSD PNC light shift in the case of the
$S_{1/2}\rightarrow D_{3/2}$ transition is about a few mHz (see
Table ~\ref{tab1}) which requires the measurement of the combined PNC
light shift to less than $0.5\%$ precision. The allowed spin flip
transitions in these isotopes are associated with not only the
desired differential PNC light shift but also a much larger
differential quadrupole light shift. Thus to achieve a precision
below $0.5\%$, all the sources of temporal variation of the
quadrupole light shift need to be stable with uncertainty well below a percent. Only
in such cases the measurements of the quadrupole and PNC
light shifts within a short time interval can provide NSD PNC light
shifts with the desired accuracies. However, a suitable choice of the hyperfine states and
the Zeeman sublevels allows to avoid the systematics from the quadrupole
light shift. In the $F=2 (S_{1/2}) \rightarrow F'=3 (D_{3/2})$
transition in spin $I=3/2$ isotopes, the sublevels $m_{F}=1,0$ of
$F=2 (S_{1/2})$ will have the same quadrupole light shift and this will reduce the
systematic error in the NSD PNC measurement. An unambiguous measurement
of NSD PNC will indeed be a challenge, since it is necessary to know
the NSI PNC part in the same isotope. An alternative is to consider
the $F=3 (S_{1/2}) \rightarrow F'=2 (D_{3/2})$ transitions in
$^{139}$Ba$^{+}$~($I=7/2$) and $^{229}$Ra$^{+}$~($I=5/2$). In these
isotopes $m_{F}=3,1$ sublevels ($F=3, S_{1/2}$) experience the same
quadrupole light shift while $m_{F}=2$ is free from the quadrupole
and PNC light shifts. Thus it would be possible to extract both
contributions to the PNC by driving the spin flip transitions after
preparing the ion in the $m_{F}=2$ state.

In principle an unambiguous measurement of NSD PNC
would be possible in the $S_{1/2}\rightarrow D_{5/2}$ transition, but the
size of the observable light shift is very small; hence its realization
is a challange. In Table~\ref{tab1} the light shifts in the $S_{1/2}\rightarrow D_{5/2}$ transitions with different hyperfine states
are given for various isotopes of these ions where it is
feasible to drive spin flip transition between the magnetic
sublevels avoiding the differential quadrupole light shift. It shows
that an experimental uncertainty below $0.05\%$ of the Larmor frequency
shift measurement is essential to extract NAM result.

In conclusion, we have given the estimated values of the light
shifts due to the nuclear spin independent and dependent parity nonconserving
interactions for various isotopes of singly ionized barium and radium isotopes.
We have shown that the size of these effects would be rather small for 
Ba$^{+}$, and 
therefore it would be quite challenging to observe the nuclear spin dependent
parity nonconserving effect 
in this ion. But in the case of Ra$^{+}$,
the prospects for observing the parity nonconserving
light-shift due to the nuclear spin dependent 
interaction are much better. An unambiguous observation of light-shift due to
nuclear spin-dependent interaction 
in the $7s \ ^2S_{1/2} \rightarrow 6d \ ^2D_{5/2}$
transition might be feasible in singly ionized radium. Our
analysis highlights the isotopes of Ra$^{+}$ and the transitions 
in them that could be suitable for observing
the nuclear spin-dependent parity nonconservation.

We are grateful to B. P. Das for his invaluable suggestions and J. Bhatt 
for critical remarks in this work. MM and PM thank the DST-SERC for 
supporting the project. Computations were carried out in PRL HPC 3TFLOP 
cluster and ParamPadma, C-DAC, Bangalore.


\begin{thebibliography}{22}
\bibitem{ginges}
J. S. M. Ginges and V. V. Flambaum, Phys. Rep. {\bf 637}, 63 (2004).
\bibitem{wood}
C. S. Wood {\it et al}, Science {\bf 275}, 1759 (1997).
\bibitem{haxton}
W. C. Haxton, C.-P. Liu and M. J. Ramsey-Musolf, Phys. Rev. Lett. {\bf 86}, 5247 (2001); Phys. Rev. C 65, 045502 (2002).
\bibitem{budker}
K. Tsigutkin {\it et al}, Phys. Rev. Lett. {\bf 103}, 071601 (2009).
\bibitem{fortson}
N. Fortson, Phys. Rev. Letts. {\bf 70}, 2383 (1993).
\bibitem{wansbeek}
L. W. Wansbeek {\it et al}, Phys. Rev. A {\bf 78}, 050501(R) (2008).
\bibitem{versolato}
O. O. Versolato {\it et al} , Phys. Rev. A {\bf 82}, 010501(R) (2010).
\bibitem{sherman}
J. A. Sherman {\it et al}, Phys. Rev. Letts. {\bf 94}, 243001 (2005).
\bibitem{koerber}
T. W. Koerber {\it et al}, Phys. Rev. Letts. {\bf 88}, 143002 (2002).
\bibitem{sahoo1}
B. K. Sahoo, R. K. Chaudhuri, B. P. Das, and D. Mukherjee, Phys. Rev. Lett. {\bf 96}, 163003 (2006).
\bibitem{sahoo2}
B. K. Sahoo, B. P. Das, R. K. Chaudhuri, and D. Mukherjee, Phys. Rev. A {\bf 75}, 032507 (2007).
\bibitem{sahoo3}
B. K. Sahoo {\it et al}, Phys. Rev. A {\bf 76}, 040504(R) (2007).
\bibitem{geetha}
K. P. Geetha, A. D. Singh, B. P. Das, and C. S. Unnikrishnan, Phys. Rev. A {\bf 58}, R16 (1998).
\bibitem{lindgren}
I. Lindgen and J. Morrison, {\it Atomic Many-Body Theory}, edited by G. Ecker, P. Lambropoulos, and H. Walther ( Springer-Verlag, Berlin, 1985).

\end{thebibliography}
\end{document}